\documentclass[aps,prl,amsmath,amssymb,reprint,superscriptaddress]{revtex4-1}
\usepackage{graphicx}
\usepackage{amsmath}
\usepackage{dcolumn}
\usepackage{bm}
\usepackage{bbold}

\bibstyle{apsrev4-1}

\begin{document}


\title{Four Wave Mixing from Fe$^{3+}$ Spins in Sapphire}

\author{Daniel L. Creedon}
\email{creedon@physics.uwa.edu.au}
\affiliation{ARC Centre of Excellence for Engineered Quantum Systems, University of Western Australia, 35 Stirling Highway, Crawley WA 6009, Australia}

\author{Karim Benmessa\"i}
\affiliation{ARC Centre of Excellence for Engineered Quantum Systems, University of Western Australia, 35 Stirling Highway, Crawley WA 6009, Australia}

\author{Warwick P. Bowen}
\affiliation{ARC Centre of Excellence for Engineered Quantum Systems, University of Queensland, St. Lucia QLD 4072, Australia}

\author{Michael E. Tobar}
\affiliation{ARC Centre of Excellence for Engineered Quantum Systems, University of Western Australia, 35 Stirling Highway, Crawley WA 6009, Australia}

\date{\today}


\begin{abstract}
Fe$^{3+}$ ions in sapphire exhibit an Electron Spin Resonance (ESR) which interacts strongly with high-$Q$ Whispering Gallery (WG) modes at microwave frequencies. We report the first observation of a third-order paramagnetic nonlinear susceptibility in such a resonator at cryogenic temperatures, and the first demonstration of four wave mixing (FWM) using this parametric nonlinearity. This observation of an all-microwave nonlinearity is an enabling step towards a host of quantum measurement and control applications which utilize spins in solids.
\end{abstract}

\maketitle

Since the development of the first laser \cite{MaimanLaser}, a multitude of nonlinear effects have been observed in optical systems. Optical second- \cite{2ndHarmonicGeneration} and third-harmonic generation \cite{3rdHarmonicGeneration,Vahala3rdHarmonic}, optical sum-frequency generation \cite{OpticalSumDifference}, optical parametric oscillation and amplification\cite{OpticalParaAmp,KippenbergVahalaPRL}, Raman lasing \cite{KippenbergRamanLaser}, and two-photon absorption \cite{TwoPhotonExcitation} are all well-characterised nonlinear effects which have been instrumental in the development of the past few decades of modern optics. High quality optical cavities allow the effect of the nonlinearity to be greatly enhanced, and have lead to many new applications including the implementations of frequency combs through parametric frequency conversion effects  \cite{KippenbergScienceComb,KippenbergNPhotComb,KippenbergNatureComb}. Optical nonlinearities are crucial for switching and modulation in modern communications technology, and are an enabling capability for future implementations of optical computer technologies, including the possibility of a quantum computer based on encoded single photons \cite{OBrien07122007}.
Recently, dramatic progress has been made in using microwave systems for quantum information and measurement, with nonlinearities playing a critical role. Josephson junctions in particular, which operate at microwave frequencies, act as a nonlinear inductor which permits uneven spacing of energy levels, leading to individual addressability of energy states using an external field. This, and other strongly nonlinear systems are currently of considerable interest for a new generation of quantum measurement experiments including quantum-limited amplification \cite{Laflamme2011}, single quadrature squeezing with tunable nonlinear Josephson metamaterials \cite{Castellanos2008}, readout of superconducting flux qubits \cite{PhysRevLett.96.127003}, and frequency conversion with quantum-limited efficiency \cite{Ramirez2011}. An addressable quantum memory with coherence times long enough for quantum computing applications could potentially be achieved through the manipulation of electron spins in a crystal lattice host, which typically occurs at microwave frequencies, and can have characteristic relaxation times of order seconds.  This, along with the potential for large collective couplings, have provoked great interest in electron spins in solids as potential quantum memories for superconducting qubits. In particular, nitrogen-vacancy (NV) centers in diamond \cite{PhysRevLett.105.140502}, Cr$^{3+}$ spins in sapphire \cite{PhysRevLett.105.140501}, and nitrogen spins in fullerene cages \& phosphorous donors in silicon \cite{PhysRevLett.105.140503} have been well studied in circuit QED experiments coupling superconducting resonators to electron spin ensembles.

\begin{figure*}[t]
\includegraphics[width=7in]{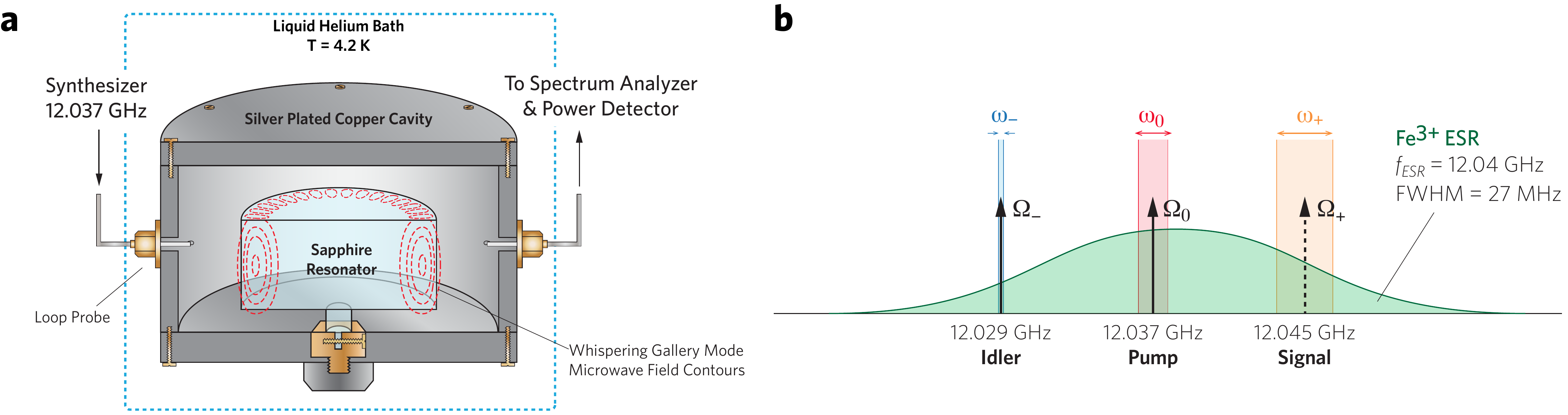}
\caption{\label{7sec-schematic} \textbf{(a)} Cross section of the experimental package. The sapphire resonator is shown in its cavity, which is mounted in a vacuum chamber at the end of an insert in a liquid helium dewar. Herotek DT8016 power detectors are used at the output to generate a voltage proportional to incident microwave power, measured with an oscilloscope.  \textbf{(b)}  (Not to scale.) Schematic of the system described by our theoretical model. $\Omega_0$ and $\Omega_{-}$ represent fixed microwave WG mode resonances in the sapphire resonator with bandwidths of order 10 Hz. $\Omega_{+}$ models a lossy resonance at $\Omega_{+} = 2\Omega_0 - \Omega_{-}$ as no WG mode exists at 12.045 GHz. The applied pump frequency $\omega_{0}$ can be selected in a range of over 4 kHz around $\Omega_0$ to successfully result in the generation of $\omega_{-}$ whose frequency only changes over a narrow range $<$40 Hz, and $\omega_{+}$ with a frequency range of order 8 kHz.}
\end{figure*}

In this Letter, we demonstrate the resonant enhancement of the weak nonlinear $\chi^{(3)}$ paramagnetic susceptibility present in a parts-per-billion concentration of electron spins in sapphire. To the authors' knowledge, this is the first observation of a paramagnetic nonlinear process purely at microwave frequencies in a crystalline host. Degenerate FWM is achieved with the application of only a single pump field, with the pump and idler frequencies enhanced by ultra-high {$Q$-factor} WG mode resonances. FWM is an enabling process for both frequency comb generation and many quantum computing and metrology applications. Our system is further suited to these applications due to the extremely low dielectric loss tangent at millikelvin temperature, which persists even at single photon input power \cite{Creedon2011}.  
\begin{figure*}[!th]
\includegraphics[width=7in]{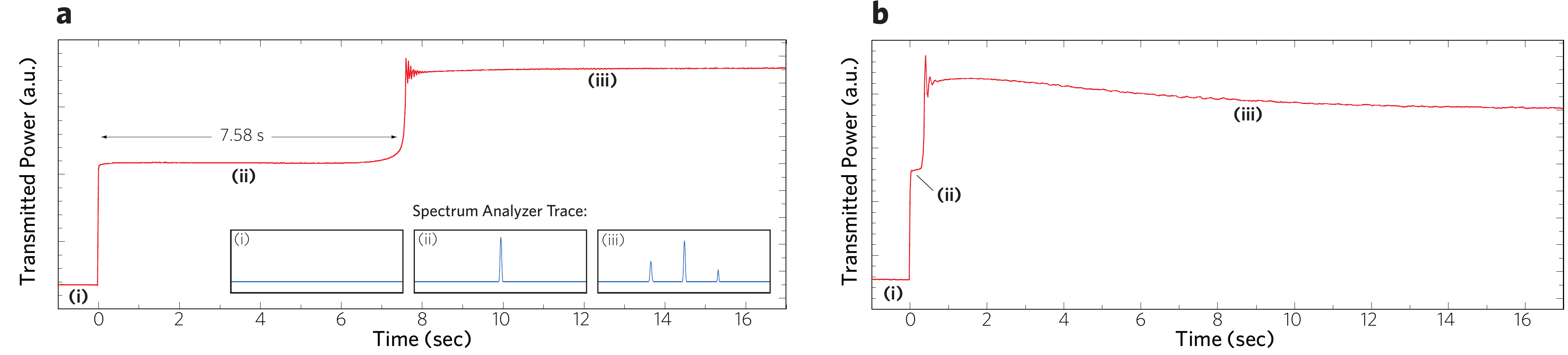}
\caption{\label{7sec}Power detected in transmission through the resonator when (a) the 12.037 GHz pump signal is offset 2.671 kHz above resonance, and (b) when the 12.037 GHz pump is offset 2.659 kHz above resonance. In both cases, the excitation signal was switched on at zero seconds.}
\end{figure*}

The experimental system consists of a cryogenic sapphire resonator-oscillator \cite{Pyb2005apl,Benmessai2007el,Benmessai2008prl,BenmessaiAmpProc,BenmessaiGyrotropic,Creedon2010} as shown in Fig. \ref{7sec-schematic}(a). The system is cooled to liquid helium temperature and pumped with microwave power to excite WG mode resonances. As a result of the manufacturing process, paramagnetic Fe$^{3+}$ are included in the sapphire lattice at a concentration of 150 ppb ($\sim10^{16}$ spins in the lattice) \cite{Creedon2010}. The crystal field splitting results in an  inhomogeneously broadened electron spin resonance (ESR) with 27 MHz linewidth \cite{BogleSymmons} at zero applied DC magnetic field, corresponding to the spin-$\left|1/2\right\rangle$, $\left|3/2\right\rangle$, and $\left|5/2\right\rangle$ states of the ion. Within the system a complex interaction occurs between the microwave input field, a dilute paramagnetic Fe$^{3+}$ spin system, and $^{27}$Al lattice ions, which ultimately results in the production of signal and idler photons equally spaced in frequency characteristic of degenerate FWM. The resonator-oscillator geometry is such that two microwave resonances exist within the ESR bandwidth, as shown in Fig. \ref{7sec-schematic}(b), which act to resonantly enchance both the pump and idler fields. The pump resonance frequency of $\omega_{0}=$ 12.0375 GHz is co-incident with the $\left|1/2\right\rangle \rightarrow \left|3/2\right\rangle$ transition residing at the maximum of the ESR, and the idler resonance frequency of $\omega_{-} = 12.0298$ GHz is in the wings of the ESR. No WG mode exists within the ESR bandwidth at the signal frequency $\omega_{+}$.
 
At low pump powers, excitation was only observed at the pump frequency. However, as is characteristic of degenerate four-wave mixing, after a threshold power level is surpassed, continuous excitation of the signal and idler fields is also found to be present, with the signal and idler frequencies equally spaced $\Delta\omega=7.669$ MHz from the pump frequency, and the idler frequency clamped to its whispering gallery mode resonance. Remarkably, signal and idler excitation was observed to appear anywhere from instantaneously to $\sim$7.5 seconds after application of the pump, with the delay being strongly dependent on the pump power and its detuning  from resonance. Figure \ref{7sec} shows the transmitted power through the resonator as a function of time for two different cases, with insets showing the spectrum analyser trace before (Fig. \ref{7sec}(a)(i)) and after applying the synthesizer signal. In the first case, when the synthesizer is switched on (Fig. \ref{7sec}(a)(ii)) the transmitted power remains constant for 7.58 seconds, after which time the signal and idler appeared simultaneously (Fig. \ref{7sec}(a)(iii)) and the total output power detected was seen to rise. In contrast, in the second example the signal and idler appear significantly faster, but relax over a period of several seconds before reaching a steady-state transmitted power level.

The complex time dynamics of the interaction can be qualitatively understood as being due to the slow seeding of the parametric process due to energy transfer in the Fe$^{3+}$ spin system. Upon application of the pump field, the sub-set of Fe$^{3+}$ ions within the ESR at the pump frequency begin to absorb energy. The hyperfine lattice interaction between the individual Fe$^{3+}$ spin packets, and the $^{27}$Al nuclear spins then slowly transfers power through a cross-relaxation process from the pump mode frequency, down to the idler resonance frequency thus seeding the signal. A similar behaviour has been observed in optical systems, where Raman scattering from the pump into the signal seeds the parametric process, and significantly reduces the threshold power for observation four-wave mixing\cite{Freitas2005314,Dahan:05}. In the optical case, the Raman scattering occurs virtually instantaneously.  Here, by contrast, the time constant of the hyperfine lattice interaction can be extremely long, with previous studies with Cr$^{3+}$ doped sapphire \cite{ElectronNuclear} recording transient relaxation times on the order of 5 seconds. We attribute the complex dynamics observed over long time scales in our experiment to this fact.

\begin{figure}[b]
\includegraphics[width=3.2in]{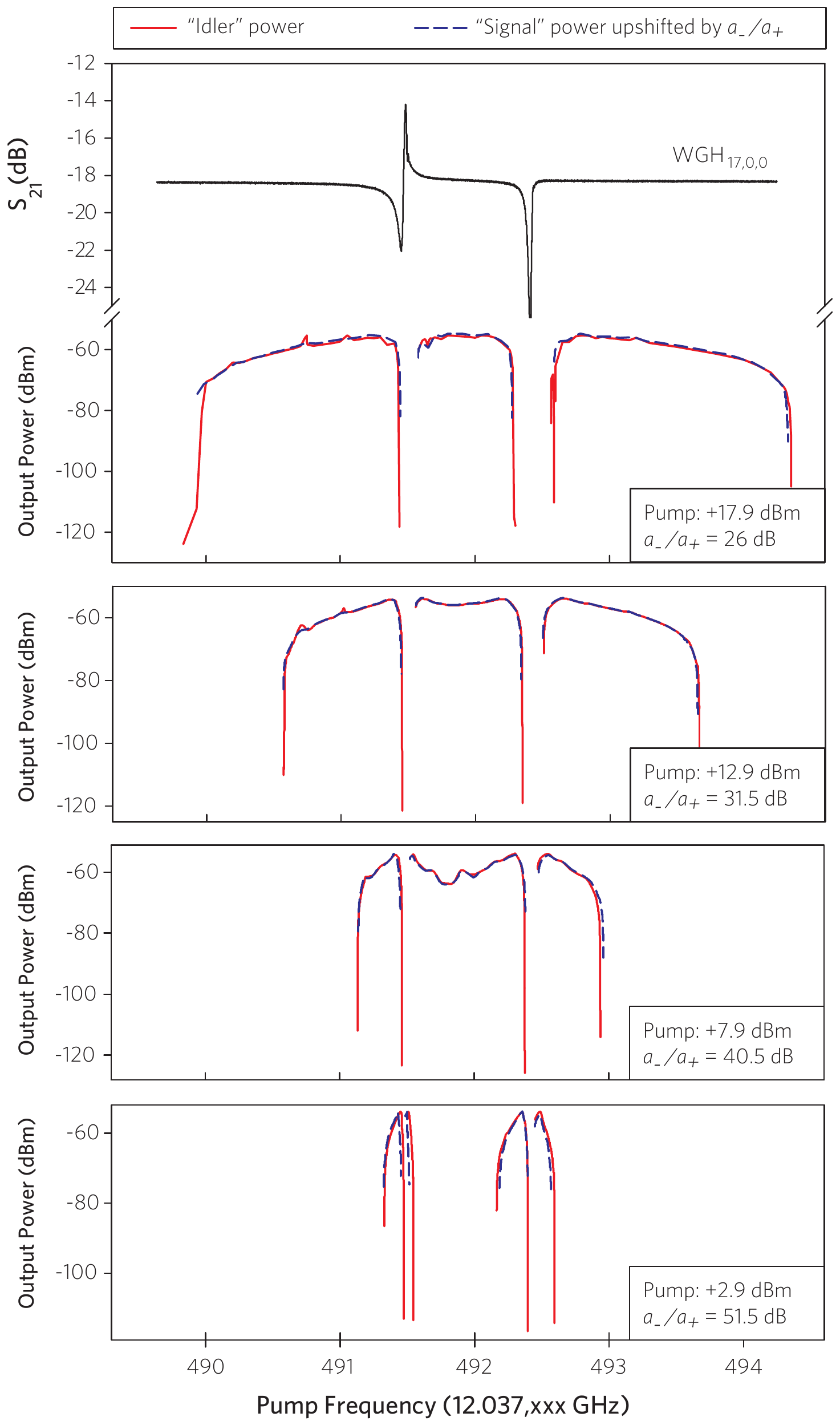}
\caption{\label{045freqpower}Output power of the signal and idler as the input pump is swept in frequency over the WGH$_{17,0,0}$ resonance.  The power of the signal is upshifted by the ratio of the amplitudes $a_{-}/a_{+}$. This upshifted curve nearly directly overlaps the signal curve. The transmission curve of the WGH$_{17,0,0}$ resonance is shown for reference.}
\end{figure}

\begin{figure}[t]
\includegraphics[width=3.3in]{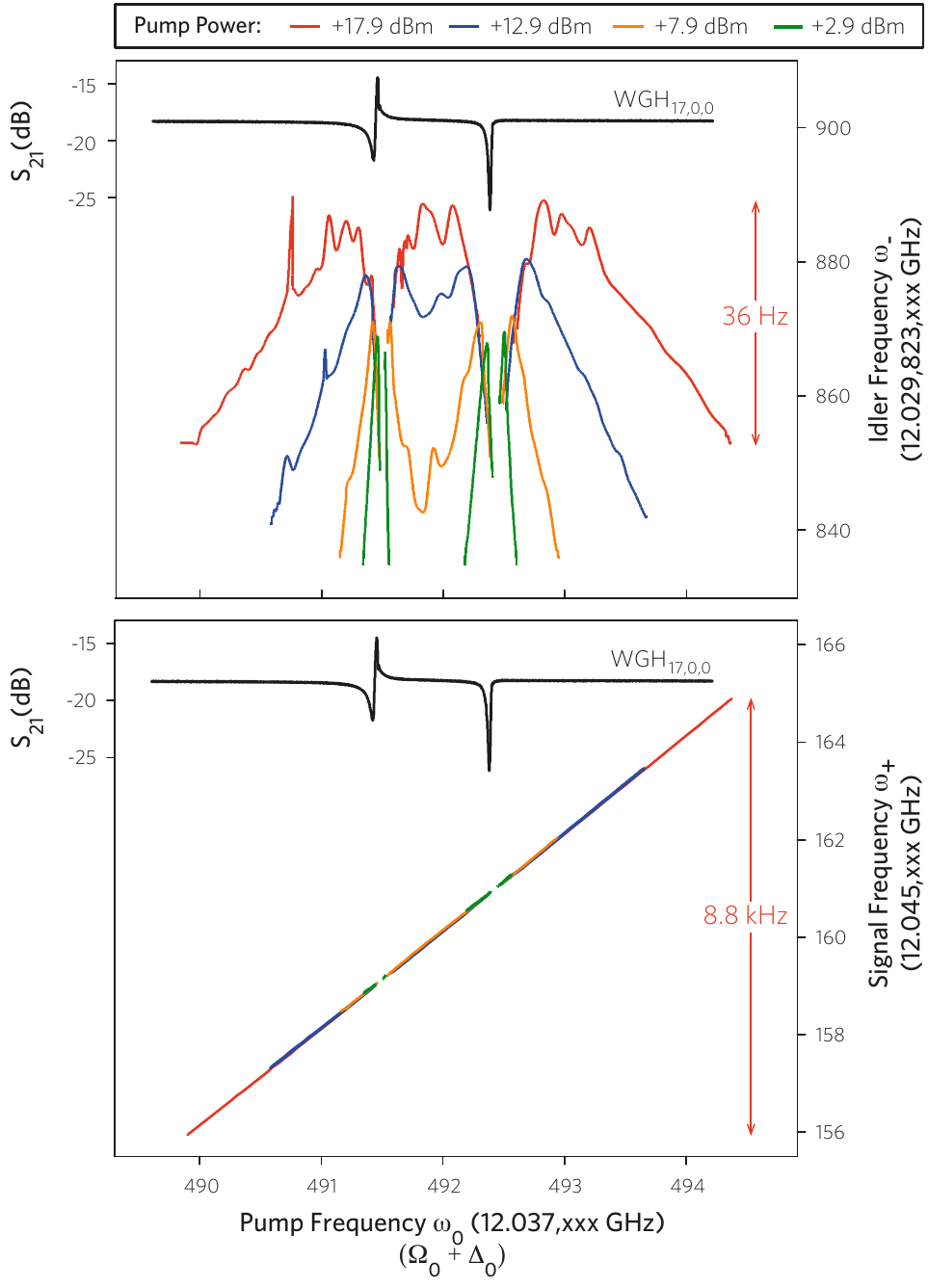}
\caption{\label{freqpull}Output frequency of the idler (top), and signal (bottom) as the input pump frequency is swept over the WGH$_{17,0,0}$ resonance at 12.037 GHz. The gradient of the `signal' slope is 2.00, meaning $\Delta_{+} \approx 2\Delta_0$. The frequency of the idler ($\omega_{-}$) is strongly locked to the WG mode frequency $\Omega_{-}$ as the frequency shift is more than two orders of magnitude smaller.}
\end{figure}

The four-wave mixing operates for pump frequencies over a range spanning 4 kHz, corresponding to $\sim$400 times the linewidth of the pump WG resonance. Figure~\ref{045freqpower} shows the output power of the signal and idler frequencies for a selection of input powers swept over the pump WG mode resonance frequency.  Four-wave mixing is a phase sensitive process, and we observe an apparent strong phase-mismatch when the pump frequency tunes closely to pump WG mode resonance. A large enough phase mismatch ensures that four-wave mixing is effectively suppressed. Due to the resonant enhancement of the idler field, it's frequency is clamped strongly to the signal resonance frequency, varying by less than 36 Hz over the full pump frequency range as shown in Fig.~\ref{freqpull}. This allows the signal frequency to be widely and predictably tuned by tuning the pump frequency. The tunable bandwidth decreases with pump power and is, for example, only several hundred Hertz at a pump power of 5 dBm.

The full model of the parametric process including cross-relaxation induced seeding is beyond the scope of this article. Here we instead neglect the seeding process, and estimate the parametric nonlinearity through a simple three mode picture with Hamiltonian:
\begin{multline}
\label{hamiltonian}
H = \hbar \Omega_0 {\hat a}_0^\dagger {\hat a}_0 + \hbar \Omega_{-} {\hat a}_{-}^\dagger {\hat a}_{-} + \hbar \Omega_{+} {\hat a}_{+}^\dagger {\hat a}_{+} \\ + i \hbar g {\hat a}_0^2 {\hat a}_{-}^\dagger {\hat a}_{+}^\dagger  -  i \hbar g^* {\hat a}_0^{\dagger 2} {\hat a}_{-} {\hat a}_{+},
\end{multline}
where the terms on the first line are the rest energy of the system, while those on the second account for the nonlinear interaction, with $g$ being the nonlinear interaction strength. The annihilation operator ${\hat a}_j$ describes the field amplitude in mode $j$, with $\langle a_j^\dagger a_j \rangle $ being the mean photon number in the mode and the subscripts $0$, $+$, and $-$ respectively denoting the pump, signal and idler modes. $\Omega_j$ is the resonance frequency of mode $j$. The splitting of the WGH$_{17,0,0}$ mode at $\Omega_{0}$ is neglected since the splitting frequency is far smaller than the spacing of the pump, signal, and idler frequencies, and thus it is expected not to contribute significantly to the physics.

\begin{figure}[h]
\includegraphics[width=3.3in]{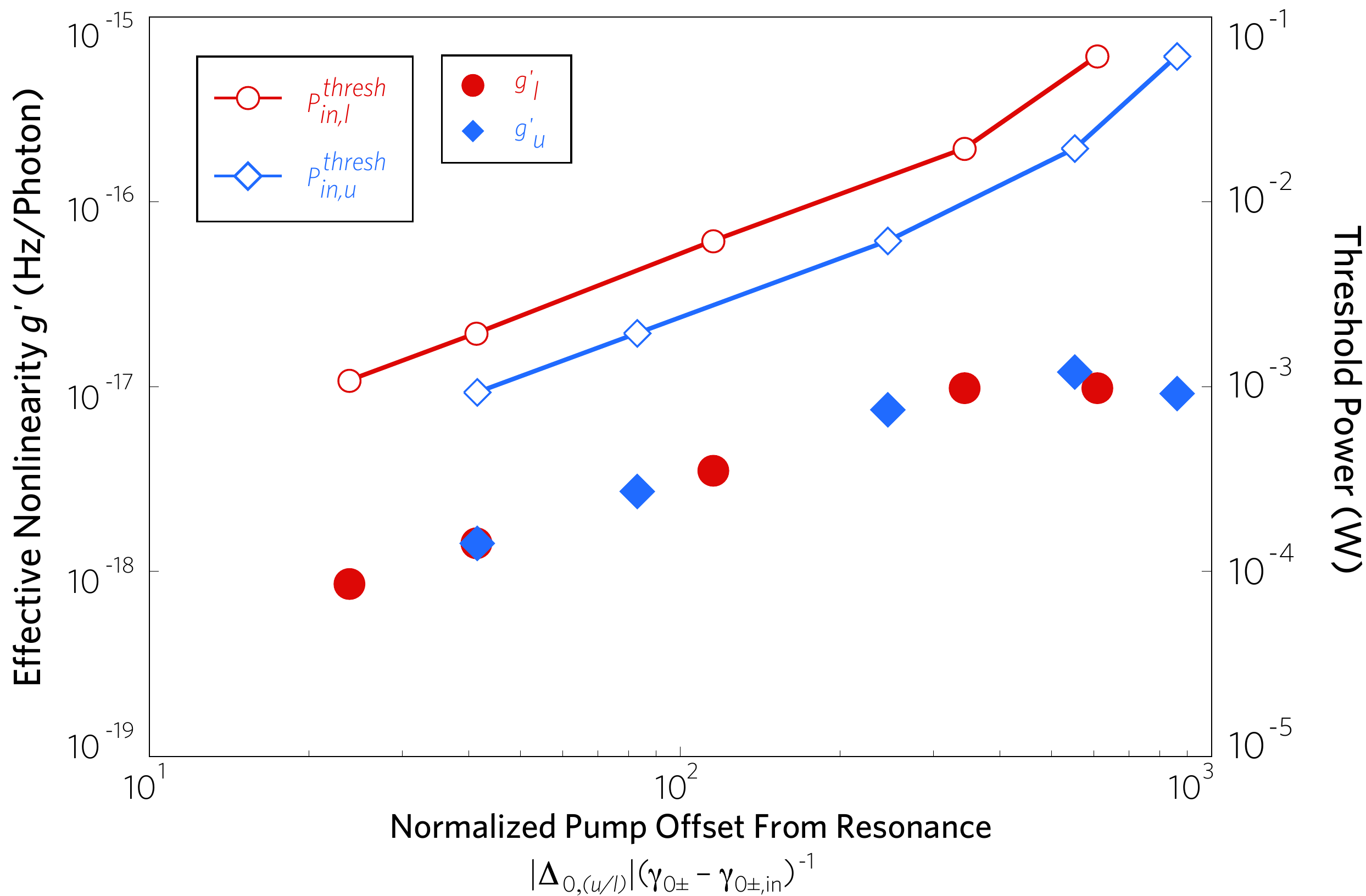}
\caption{\label{nonlin}Effective nonlinearity and threshold power as a function of the normalised detuning from resonance. The subscripts $u$ and $l$ refer respectively to the upper doublet of the WGH$_{17,0,0}$ mode (for positive detuning), and lower doublet (for negative detuning). Here, $g'$ ranges between 10$^{-18}$ and 10$^{-17}$, and decreases as the pump tunes towards resonance, with FWM turning off between 20 and 40 unloaded half bandwidths (HBWs) from resonance. A maximum in $g'$ occurs around 500 unloaded HBWs, with an associated inflection in the required $P_{\rm in}^{\rm thresh}$ to enable FWM.}
\end{figure}

Applying the quantum Langevin Equation to Eqn. \ref{hamiltonian} \cite{McRae}, one can then find equations of motion for the pump, idler and signal. In the rotating frame, this yields the expectation value equations
\begin{eqnarray}
\dot{\alpha_0} &=& -2g\alpha_0^* \alpha_{-} \alpha_{+} - (\gamma_0 + i\Delta_0) \alpha_0 - \sqrt{2 \gamma_{0, \rm in}} \alpha_{0, \rm in} \label{pump}\\
\dot{\alpha_-} &=& g \alpha_0^2 \alpha_+^*- (\gamma_- + i \Delta_-) \alpha_- \label{sideband1}\\
\dot{\alpha_+} &=& g \alpha_0^2 \alpha_-^* - (\gamma_+ + i \Delta_+) \alpha_+ \label{sideband2}.
\end{eqnarray}
where $\alpha_j = \langle a_j \rangle$, $\gamma_j$ and $\Delta_j$ are, respectively, the decay rate and detuning from resonance of field $j$, $\gamma_{0,\rm in}$ and $\alpha_{0,\rm in}$ are the input coupling and amplitude of the incident pump field, and since the idler and signal are not pumped, $\alpha_{-, \rm in} = \alpha_{+,\rm in}=0$. Expressed in terms of half-bandwidths, $\gamma_{-}=$ 6 Hz, $\gamma_{0,l}=$ 5 Hz, and $\gamma_{0,u}=$ 6.7 Hz.  Consistent with our experiments, we model the signal resonance as a lossy resonance such that the signal dynamics are fast compared with the pump and idler dynamics. Hence, $\gamma_{+}$ represents a lossy damping and is related to the ratio of the amplitudes $\gamma_{+} = \gamma_{-} \frac{a_{-}^2}{a_{+}^2}$. Equation~(\ref{sideband2}) can then be adiabatically eliminated, giving an equation of motion for the signal:
\begin{multline}
\dot{\alpha_{-}} = - \left[ \gamma_{-}\left(1- g' |\alpha_0|^4 \right) \right. \\ \left. + i \left(\Delta_1 \left(1 -  g' \frac{\Delta_2 \gamma_{-}}{\Delta_1 \gamma_{+}}|\alpha_0|^4 \right)\right) \right] \alpha_{-}
\end{multline}
%
%
The effective nonlinearity $g'$ can be related to the intrinsic nonlinearity $g$, and expressed in terms of only the pump parameters, given by:
\begin{eqnarray}
g'	&=& g \frac{\gamma_{+}}{\gamma_{-}}  \frac{1}{\sqrt{\gamma_{+}^2+\Delta_{2}^2}} \\
	&=&  \frac{\gamma_0^2 + \Delta_0^2}{2 \gamma_{0, \rm in}}  \frac{  \hbar \Omega_0}{P_{\rm in}^{\rm thresh}}
\end{eqnarray}
%
%
Here, the steady-state intracavity pump amplitude $\alpha_0$ is related to the threshold power by $P_{\rm in}^{\rm thresh} =  \hbar \Omega_0 | \alpha_{0,\rm in}^{\rm thresh}|^2$. Figure \ref{nonlin} shows the effective nonlinearity and threshold power as a function of normalised detuning from resonance, calculated using the parameters of our system.

In summary, we have demonstrated that a strong $\chi^{(3)}$ nonlinearity at microwave frequencies, arising from only a parts-per-billion concentration of paramagnetic ions, leads to degenerate four-wave mixing in a cryogenic sapphire resonator when pumped with only a single frequency. Long characteristic times on the order of several seconds were observed due to slow cross-relaxation and interaction with lattice ion nuclear spins, and broad tunability can be achieved by altering the pump frequency. Our system has the potential for application in a host of future quantum computing and metrology experiments where low microwave loss and strong nonlinearity is desirable, such as measurement of qubits in circuit QED setups, single quadrature amplification and squeezing, quantum-limited parametric amplification, or potential use in nanoscale magnetometry \cite{PhysRevB.83.134501} with the benefit of having the necessary amplification and nonlinearity integrated within a single device.\\

\begin{acknowledgments}
This work was funded by Australian Research Council grant numbers FL0992016 and CE11E0082.
\end{acknowledgments}

\bibliography{biblio}

\end{document}